\documentclass[11pt,fleqn]{article}
\usepackage{graphicx}
\usepackage{epsfig}
\newcommand{\Lnoise}[1]{{\cal L}_\sigma^{#1}}	% noisy evolution operator
\newcommand{\Prpgtr}[1]{\Delta_{#1}}
\newcommand{\InvPrpgtr}[1]{\Delta^{-1}_{#1}}
\newcommand{\CrlMat}[1]{C_{#1}}

\newcommand{\field}{\phi}
\newcommand{\Df}[1]{f^{'}_{#1}}
\newcommand{\mod}[1]{\dot{#1}}
\newcommand{\const}[1]{\hat{#1}}

\newcommand{\btrack}[1]{\raisebox{-2.0ex}[3.5ex][2.5ex]
	{\includegraphics[height=5ex]{figs/f_#1.eps}\negthinspace} }
\newcommand{\btrackB}[1]{\raisebox{-4.0ex}[5.5ex][4.5ex]
          { \epsfig{file=figs/f_#1.eps,height=9ex}\negthinspace} }
\newcommand{\btrackBB}[1]{\raisebox{-5.0ex}[6.5ex][5.5ex]
          { \epsfig{file=figs/f_#1.eps,height=11ex}\negthinspace} } 
\newcommand{\rf}     [1] {~\cite{#1}}
\newcommand{\refref} [1] {ref.~\cite{#1}}
\newcommand{\refrefs}[1] {refs.~\cite{#1}}

\newcommand{\refeq}  [1] {(\ref{#1})}

\newcommand{\reffig} [1] {fig.~\ref{#1}}

\newcommand{\reftab} [1] {table~\ref{#1}}

\newcommand{\refsect}[1] {sect.~\ref{#1}}

\newcommand{\refappe}[1] {appendix~\ref{#1}}

\newcommand{\beq}{\begin{equation}}
\newcommand{\continue}{\nonumber \\ }
\newcommand{\nnu}{\nonumber}
\newcommand{\eeq}{\end{equation}}
\newcommand{\ee}[1] {\label{#1} \end{equation}}
\newcommand{\bea}{\begin{eqnarray}}
\newcommand{\ceq}{\nonumber \\ & & }
\newcommand{\eea}{\end{eqnarray}}
\newcommand{\barr}{\begin{array}}
\newcommand{\earr}{\end{array}}

\newcommand{\FIG}[4]{\begin{figure}
		      \centering{#1}
                      \caption[#2]{#3}
                      \label{#4} \end{figure} }

\newcommand{\evOper}{evolution oper\-ator}

\newcommand{\FPoper}{Perron-Frobenius oper\-ator} % Pesin's ordering

\newcommand{\dzeta}{dyn\-am\-ic\-al zeta func\-tion}

\newcommand{\Fd}{spec\-tral det\-er\-min\-ant}
\newcommand{\fd}{spec\-tral det\-er\-min\-ant}

\newcommand{\obser}{a}		% an observable from phase space to R^n
\renewcommand{\det}{\mbox{\rm det}}

\newcommand{\tr}{{\rm tr}\, }

\newcommand{\Lop}{{\cal L}}	   % evolution operator
\newcommand{\ExpaEig}{\Lambda}	   
\newcommand{\eigenvL}{{\nu}}       %This is really eigenvalue, not log!
\newcommand{\inFix}[1]{{\in \mbox{\footnotesize Fix}f^{#1}}}

\newcommand{\derJ}{{d~\over dJ}}

\newcommand{\cl}[1]{{n_{#1}}}	% discrete length of a cycle, Predrag
\thicklines
\newlength{\Fsize}   % allow for easy resizing of diagrams
\newlength{\Fdotsize}
\title{
	 Trace formulas for stochastic evolution operators:\\
	 Weak noise perturbation theory
      }

\author{
	Predrag Cvitanovi\'c,
	C.P. Dettmann,
	Ronnie Mainieri, and G\'abor Vattay
\\
	{\em Center for Chaos and Turbulence Studies}\\
	Niels Bohr Institute\\
	Blegdamsvej 17, DK-2100 Copenhagen \O\
	}

\begin{document}
%\doublespace
\maketitle
\\
{\bf keywords:} noise, stochastic dynamics, cycle expansions, semiclassical limit, periodic orbits, Perron-Frobenius operator, spectral determinant, zeta functions.

%DS \newpage

\begin{abstract}
%%%%
%\file{alf.nbi.dk:predrag/articles/noise/noise.tex		\\
%		20th draft, (first  draft July 93)
%              	23/7-98  ~~~-~~~ printed \today
%}
%%%%%

Periodic orbit theory is an effective tool for the analysis of classical
and quantum chaotic systems.  In this paper we extend this
approach to stochastic systems, in particular to
mappings with additive noise.  The theory is cast in the standard field
theoretic formalism, and weak noise perturbation theory written in terms of
Feynman diagrams.  The result is a stochastic analog of the next-to-leading
$\hbar$ corrections to the Gutzwiller
trace formula, with long time averages calculated from
periodic orbits of the deterministic system.  The perturbative corrections
are computed analytically and tested numerically on a simple 1-dimensional
system.

\end{abstract}

\section{Introduction}

Noise plays important role in a variety of physical contexts.
Robustness to noise is of interest for any system since there is always
some small length scale at which the dynamics is affected by thermal
or quantum fluctuations or unobserved degrees of freedom.
For example, the interplay of deterministic dynamics and
magnetic diffusivity is subject of great interest in the
dynamo problem, where the effect of magnetic field diffusion
on the steady fast kinematic dynamo rates is discussed in\rf{CO94}
within the periodic orbit theory formulation of \refrefs{AG93,dynamo,CG96}.

The noise
tends to regularize the theory, replacing the deterministic delta function
evolution operators by smooth distributions.
While in this paper we are interested in effects of weak but
{\em finite} noise,
the $\sigma \to 0$ limit is also important as a tool for identifying
the natural measure\rf{sinai,bowen,ruelle} for deterministic flows.
The noise regularization might in addition cure some of the ills of
intermittent systems which are plagued by power-law convergences
arising from terms like $|\Lambda-1|^{-1}$
in the limit $\Lambda \to 1$.

We have cast the theory in the standard field theoretic
language\rf{FieldThe}, in the spirit of approaches such as the
Martin-Siggia-Rose\rf{msr} formalism, the Parisi-Wu\rf{PW}
stochastic quantization, and the
Feigenbaum and  Hasslacher\rf{FH} study of noise renormalization
in period doubling.
This perturbation theory has the same structure as
the $\hbar$ corrections
to the semiclassical Gutzwiller trace formulas\rf{gutbook}
computed by Gaspard and Alonso\rf{alonso1,alonso2,gasp_hbar},
and the trace formulas for continuous stochastic
flows and for the $\hbar$ corrections formulated by
Vattay\rf{vattay_BS}.

Though it is clear from the literature on stochastic path integrals
that some kind of Feynman diagrams apply, the present work seems to
be one of the few that actually compute the weak noise corrections
for a concrete dynamical system, although in some cases the leading
correction may be obtained directly from the perturbed
eigenfunction\rf{reimann1,reimann2}.
The form of the perturbative expansions  of \refsect{s:WeakNsPertExp}
is reminiscent of perturbative calculations
of field thery, but in some aspects the 
calculations undertaken here are relatively more difficult. 
The main difference is that there is
no translational invariance along the chain, so unlike the case of
usual field theory,
the propagator is not diagonalized by a Fourier transform. We
do our computations in configuration coordinates.
Unlike the most field-theoretic literature,
we are neither ``quantizing'' around a trivial vacuum,
nor a countable infinity of stable soliton saddles, but around an
infinity of nontrivial unstable hyperbolic saddles.

Two aspects of our results are {\em a priori} far from obvious:
(a) that the structure of the periodic orbit theory
should survive introduction of noise, and (b) 
a more subtle and surprising result,
repeats of prime cycles can be resummed and theory reduced to the
\dzeta s and \fd s of the same form as the for the deterministic
systems.

Having constructed the perturbation expansion in \refsect{s:WeakNsPertExp},
in \refsect{e:NumTests} we confront the theory
with a numerical determination of 
eigenfunctions and eigenvalues, and verify the
correctness of our perturbation expansion to the same 
numerical accuracy.
A variety of flow models with noise are simpler to study in nonperturbative
(large $\sigma$) limits; numerical eigenfunctions do not
depend on the weak noise assumptions, and in fact require the noise
to be larger than the effective discretization length of the basis.

\section{Stochastic evolution operator}

The periodic orbit theory allows us to calculate long time averages
in a chaotic system as expansions in terms of the periodic orbits
(cycles) of the system.  
The simplest example is provided by the {\FPoper} 
\[
\Lop \rho(y)=\int dx\,\delta(y-f(x))\rho(x)
\]
for a {\em deterministic} map $f(x)$ which maps a density distribution
$\rho(x)$ forward in time.
The periodic orbit theory relates the spectrum of this operator and
its weighted evolution operator generalizations to the periodic orbits
via trace formulas, \dzeta s and \fd s\rf{PG97,QCcourse}. Our
purpose here is to develop the parallel theory for {\em stochastic}
dynamics, given by
the discrete Langevin equation\rf{vk,LM94}
\begin{equation}
x_{n+1}=f(x_n)+\sigma\xi_n
\,,\label{Langevin}
\end{equation}
where the $\xi_n$ are independent normalized Gaussian random variables.

We shall treat a chaotic system with such Gaussian weak external noise by 
replacing the the deterministic evolution $\delta$-function kernel 
by $\Lnoise{}$,  the Fokker-Planck
kernel corresponding to (\ref{Langevin}),
a sharply peaked noise distribution function 
\beq
\Lnoise{} =\delta_\sigma(y-f(x))
\,,
\ee{Lnoise}
where  $\delta_\sigma$ is the Gaussian kernel
\beq
\delta_\sigma(z)=\frac{1}{\sqrt{2\pi\sigma^2}} e^{-z^2/2\sigma^2}
\,.
\ee{GaussKrnl}
The method can be applied to smooth distributions
other than the Gaussian one in the same manner.

We shall evaluate the trace formulas by steepest descent methods, and
obtain the noisy traces (traces of $\Lnoise{}$) and determinants in terms of the cycles of the
deterministic system.
The theory is then tested numerically on one-dimensional maps, but we expect
the generalization to higher dimensions to be of the same structure
as the formulas derived here.

\section{Stochastic trace formula, steepest descent approximation}

We start by calculating the trace of the $n$th iterate of
the stochastic evolution operator $\Lnoise{}$
for a one-dimensional analytic
map $f(x)$ with additive Gaussian noise $\sigma$.  
This trace is an $n$-dimensional
integral on $n$ points along a discrete periodic chain, 
so $x$ becomes an $n$-vector $x_a$ with indices $a,b,\ldots$
ranging from $0$ to $n$$-$$1$ 
in a cyclic fashion
\bea
\tr{\Lnoise{n}} &=& \int[dx]\, \exp\left\{-\frac{1}{2\sigma^2}
\sum_{a}\left[x_{a+1}-f(x_a)\right]^2\right\}
	\continue
x_n  &=&  x_0 \,,\qquad [dx]=\prod_{a=0}^{n-1}{dx_a \over \sqrt{2\pi\sigma^2}}
\,.
\label{IntDef}
\eea
As we are dealing with a path integral on a finite discrete chain,
we find it convenient to rewrite the exponent in matrix notation
\beq
\tr{\Lnoise{n}} =\int[dx]\, 
     e^{-\left[h^{-1}x -  f(x) \right]^2/2\sigma^2}
\,,\qquad
h_{ab}=\delta_{a,b+1}
\,,
\ee{eLnoisMtrx}
where $x$ and $f(x)$ are column vectors with components $x_a$ and $f(x_a)$
respectively,
and $h$ is the left cyclic shift or hopping matrix satisfying
$h^n=1$, $h^{-1}=h^{T}$.
Unless stated otherwise, we shall assume the repeated
index summation convention throughout, and that the
Kronecker $\delta$ function is the periodic one, defined by
\beq
\delta_{ab} = {1 \over n} \sum_{k=0}^{n-1} e^{i2\pi (a-b)k/n}
\,.
\ee{CycKronck} 

For sufficiently short chains, \refeq{IntDef} is an integral that
conceivably lends itself to numerical evaluation\cite{D98}, although
clearly not in the long time $n\to\infty$ limit.
However, if the noise is weak, the path integral \refeq{IntDef}
is dominated by periodic deterministic
trajectories.
Assuming that the periodic points of given finite period
$n$ are isolated and the trajectory broadening
$\sigma$ 
sufficiently small so that they remain clearly separated, the dominant
contributions come from neighborhoods of periodic points; 
in the
{\em saddlepoint approximation} the trace \refeq{IntDef} is given by
\beq
\tr{\Lnoise{n}} \longrightarrow \sum_{x_c\inFix{n}} e^{W_c}
\,,
\ee{SdlptSum}
where the sum goes over all periodic points $x_c = x_{c+n}$ of period $n$,
$f^n(x_c)=x_c$. The contribution of the
$x_c$ neighborhood is obtained by
shifting the origin of integration to
\[
x_a \to x_a + \field_a
\,,
\]
where from now on $x_a$ refers to the position of the $a$-th periodic
point, 
and expanding $f$ in Taylor series around each of the periodic points
in the orbit of $x_c$.

The contribution of the neighborhood
of the periodic point $x_c$ is given by
\bea
e^{W_c} &=& \int[d\field]\, 
     e^{-\left(\InvPrpgtr{}\field_{} - V'(\field) \right)^2/2\sigma^2}
	\continue
%	 &=&  \int[d\varphi][d\overline{\psi}][d\psi]
%     e^{-\overline{\psi}\left(\InvPrpgtr{} -  V''(\field) \right)\psi 
%		  -\varphi^2/2\sigma^2}
%		\continue
	&=& |\det \Prpgtr{}| \int[d\varphi]\, 
     e^{\sum{1\over k} \tr\left(\Prpgtr{}V''(\field) \right)^k}
     e^{-\varphi^2/2\sigma^2}
\label{eWcMtrx}
\eea
where the propagator and interaction terms are collected in
\beq
\InvPrpgtr{ab}\field_{b} = -\Df{}(x_{a})\field_{a}+\field_{a+1}
\,,\qquad
V(\field)= \sum_a \sum_{m=2}^{\infty}f^{(m)}(x_{a})
\frac{\field_{a}^{m+1}}{(m+1)!}
\,.
\ee{DefPrpg}
We find it convenient to also introduce a bidirectional propagator
$C=\Prpgtr{}\Prpgtr{}^T$ for reasons that will become apparent below.
In the second line of (\ref{eWcMtrx}) we have changed coordinates,
\beq
\varphi = \InvPrpgtr{}\field_{} - V'(\field)
\,,
\ee{eChaCoor}
and used the matrix identity $\ln\det M = \tr\ln M$ on the Jacobian
\beq
{1 \over \det \left(\InvPrpgtr{} - V''\right)}
 = {\det \Prpgtr{} \over
                    \det \left(1 - \Prpgtr{} V''\right)}
 = \det \Prpgtr{} \,
    e^{ -\tr \ln \left(1 - \Prpgtr{} V''\right) }
\,.
\ee{e:DetIdent}

The functional dependence
of $\field=\field(\varphi)$ is recovered by iterating (\ref{eChaCoor})
\beq
\field_a = \Prpgtr{ab}\varphi_b + \Prpgtr{ab}V_b'(\field)
\,.
\ee{eIterField}

The above manipulations are standard\rf{msr} and often
used in the ``stochastic quantization'' literature\rf{PW,DamHu},
where they are artfully employed to
promote identities such as
$\det M / \det M = 1$ to supersymmetric field theories.
Such symmetries do not seem to simplify the calculation at hand.

The saddlepoint expansion is most conveniently evaluated in terms of
Feynmann diagrams, which we now introduce.
The interaction terms in $V$ and its derivatives
can be represented in terms of the vertices
\[
f^{''}(x_a)
\,=\,
\btrack{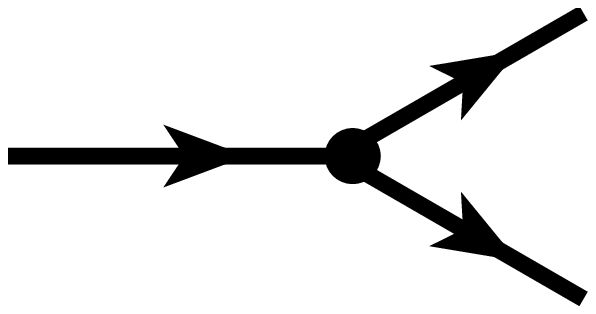}
\,,\quad
f^{'''}(x_a)
\,=\,
\btrack{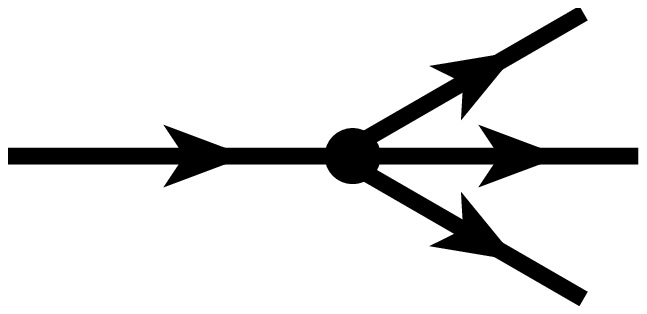}
\,,\quad
\dots
\,,
\]
\vspace{0.5\Fsize}
and the propagators as directed lines
\bea
\Prpgtr{ab}
&=&
\btrack{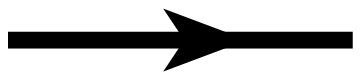}
\continue
\CrlMat{ab}
&=&
\btrack{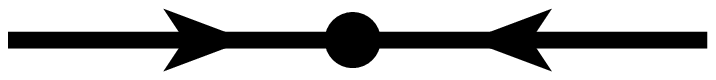}
\nnu
\eea              
The first two derivatives of $V$ may be written
\begin{equation}
\btrackBB{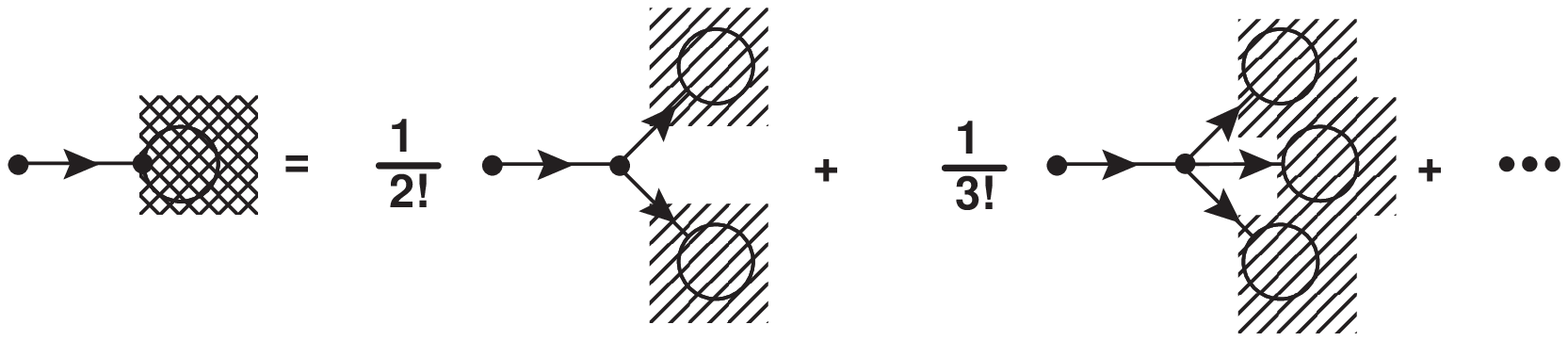} \;\;,
\end{equation}
\begin{equation}
\btrackB{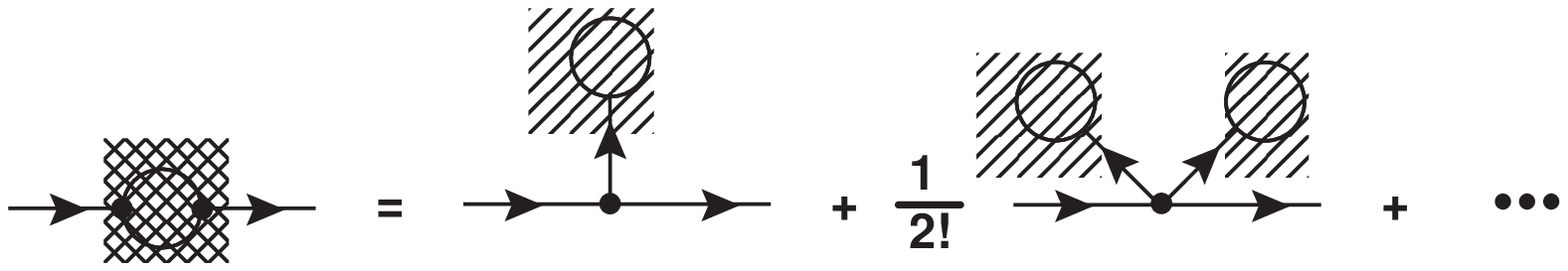}\label{e:V''}
\end{equation}
with the cross-hatched circle as $V$ and the diagonally filled
circle as $\field$.  The relation between the fields (\ref{eIterField})
becomes
\begin{equation}
\btrack{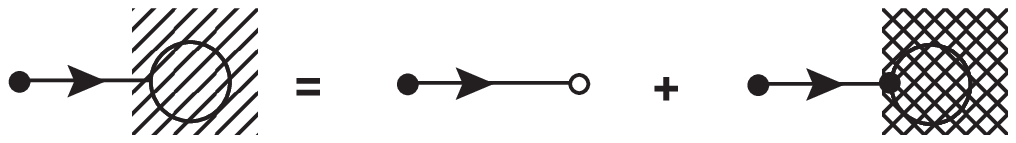}\label{e:phi}
\end{equation}
with the small open circle as $\varphi$.  This recursively
generates all tree diagrams ending in $\varphi$,
which the stochastic averaging
of \refsect{s:WeakNsPertExp} will tie into
loop corrections.

As the sum is cyclic, 
$e^{W_c}$ is the same
for all periodic points in a given cycle, independent of the choice
of the starting point $x_c$.

In the saddlepoint approximation we assume that the map is analytic
and the extrema $f^n$ are isolated.
For the leading $\sigma^2$ correction that we shall evaluate here
we need derivatives of $f$ up to the third.
A map with  non-analytic points or marginal stability would
lead to additional diffraction corrections that we shall
not consider here.

From the second path integral representation in \refeq{eWcMtrx} it follows
that $\Prpgtr{}$ can be interpreted as the ``free'' propagator. 
As $\Prpgtr{}$ will play a central role in what follows, we 
write its inverse in its full [$n$$\times$$n$] matrix form: 
\beq
\InvPrpgtr{}
	= h^{-1} - {\bf f'}
=\pmatrix{-\Df{0} &  1    &        &        &      \cr
                  & -\Df{1}&  1    &        &      \cr
                  &       & -\Df{2}&  1     &      \cr
                  &       &        & \ddots &      \cr
             1    &       &        &        & -\Df{n-1} 
         }
\ee{DeltaInv}
where ${\bf f'}$ is a diagonal matrix with elements
$ \Df{a}= \Df{}(x_a)$ a shorthand notation
for stability of the map at the periodic point $x_a$. 
The determinant of $\Prpgtr{}$ is
\beq
\det\,\Prpgtr{}={(-1)^n \over \ExpaEig_c-1}
\,, \qquad
\ExpaEig_c = \prod_{a=0}^{n-1}\Df{}(x_a)
\,,
\ee{detDel}
with $\ExpaEig_c$ the {\em stability} of the $n$ cycle going through
the periodic point $x_c$. We shall assume that we are dealing with
a chaotic dynamical system, and that all cycles are unstable,
$|\ExpaEig_c|>1$.

The formula for propagator itself is obtained by inverting \refeq{DeltaInv} 
and using relation $(h{\bf f'})^n = \ExpaEig_c$, 
(due to the periodicity of the chain):
\begin{eqnarray}
\Prpgtr{}&=& -\frac{1}{1-{\bf f'}^{-1}h^{-1}}{\bf f'}^{-1}
= -\sum_{k=0}^\infty ({\bf f'}^{-1}h^{-1})^k {\bf f'}^{-1}\nonumber\\
% &=&- {1 \over 1-\ExpaEig_c^{-1}} \sum_{k=0}^{n-1}
&=&- {1 \over \ExpaEig_c -1} \sum_{k=0}^{n-1} h ({\bf f'}h)^k
\label{InvDel}
\end{eqnarray}
In the full matrix form, the propagator is given by
\beq
\Prpgtr{} = {-1 \over \ExpaEig_c-1}
\pmatrix{ 
\Df{1}...\Df{n-1}&\Df{2}...\Df{n-1} &\Df{3}...\Df{n-1}&& \ldots&  1  \cr
   1  & \Df{2}...\Df{0}& \Df{3}\Df{4}...\Df{0} &&\ldots& \Df{0} \cr
\Df{1}&      1             & \Df{3}...\Df{0}\Df{1}&&\ldots& \Df{0}\Df{1} \cr
\Df{1}\Df{2}&\Df{2}        &     1 &\ddots&    & \Df{0}\Df{1}\Df{2} \cr
\Df{1}\Df{2}\Df{3}&\Df{2}\Df{3} & \Df{3}    &  &\ddots  &    \vdots  \cr
    \vdots       & \vdots       & \vdots&\vdots&        &    \vdots  \cr
\Df{1}...\Df{n-2}&\Df{2}...\Df{n-2} &\ldots&\ldots &1& \Df{0}...\Df{n-2}}
\label{DelMatr}
\eeq
or, more compactly,
\beq
\Prpgtr{ab} = {-1 \over \ExpaEig_c-1} \prod_{d=b+1}^{a-1}\Df{}(x_d)
\,,\qquad \Prpgtr{a,a-1}=\frac{-1}{\ExpaEig_c-1}
\,,
\label{e:Prpgtr}
\eeq
where $d$ increases cyclically through the range $b+1$ to $a-1$;
for example, if $a=0$, $a-1=n-1$.
We note that $\Prpgtr{}$ is invertible only for cycles which are
not marginal, $|\ExpaEig_c| \neq 1$. The  $|\ExpaEig_c| = 1$
case we would require going beyond the Gaussian saddlepoints
studied here, and typically to the
Airy-function type stationary points\rf{BH86}.

\section{Weak noise perturbation expansion}
\label{s:WeakNsPertExp}

The saddlepoint approximation \refeq{eWcMtrx} is a discrete path integral
on periodic chain of $n$ points which we shall evaluate by standard 
field-theoretic methods. 
Separating the quadratic terms we obtain
\beq
e^{W_c} 
	 = {1 \over |\ExpaEig_c-1|}
 	  \int[d\varphi]\,  e^{- S_0(\varphi) - S_I(\varphi)}
\,,
\ee{PropIntr1}
where
\beq
S_0(\varphi)  =  {\varphi^2}/{2\sigma^2} 
	\,, \qquad
S_I(\varphi)  = -
     \sum_{k=1}^\infty {1\over k}
\tr\left[\Prpgtr{}V''(\field(\varphi)) \right]^k
\label{PropIntrct}
\eeq
The terms collected
in $S_I(\varphi)$, linear or higher in $\varphi$, are the interaction
vertices. 

Next introduce a source term $J_a$ and define a partition function
\bea
e^{W_c(J)} &=& {1 \over |\ExpaEig_c-1|}
         \int [d\varphi]  e^{-S_0(\varphi)-S_I(\varphi) + J_a\varphi_a}
                        \continue
	   &=&
            {1 \over |\ExpaEig_c-1|} e^{-S_I(\derJ)}\int [d\varphi]  e^{-S_0(\varphi) + J_a\varphi_a}
                        \continue
	   &=&
        {1 \over |\ExpaEig_c-1|} e^{-S_I(\derJ)}  \,
           e^{ {\sigma^2 \over 2} J^2}
\,.
\label{ePartFct}
\eea
Here we have used standard formulas for Gaussian integrals
together with the normalization \refeq{IntDef}.
In our diagrammatic notation this is
\begin{equation}
\btrackB{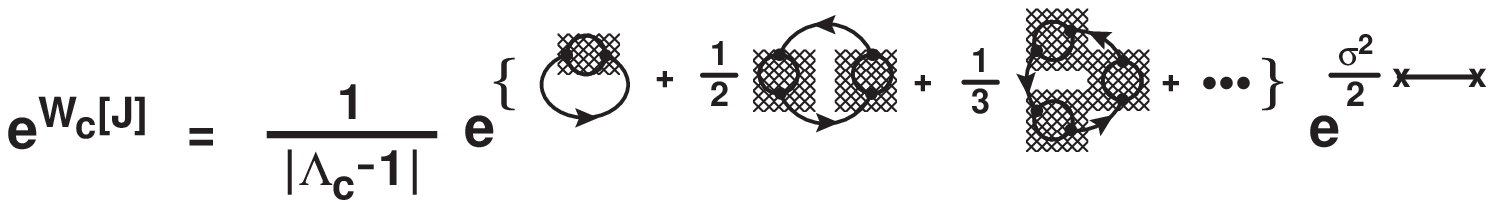}\label{e:expdiag}
\end{equation}
Expanding
\beq
e^{ {\sigma^2 \over 2} J^2}
 = 1 + {\sigma^2 \over 2} J_a J_a
     + {\sigma^4 \over 8} J_a J_a J_b J_b
     + \dots
\,,
\ee{e:Wick}
operating on this series with $\exp\{-S_I(\derJ)\}$,
\[
\left. {d \over d J_a}{d \over d J_b} 
	e^{ {\sigma^2 \over 2} J_d J_d} \right|_{J=0}
 =
               \sigma^2 \delta_{ab}
\,, \quad \cdots \,,
\]
collecting terms of the same order in $\sigma^2$, and setting $J_a$
to zero yields the perturbation expansion
\beq
W_c = - \ln|\ExpaEig_c-1| + \sum_{k=1}^\infty W_{c,2k}\sigma^{2k}
\,.
\ee{e:PertExpW}
In field-theoretic calculations the $ W_{c,0}$ term
is usually an overall volume term that drops out in the expectation
value computations. In contrast, here the
$ W_{c,0} = - \ln|\ExpaEig_c-1|$ term 
is  the classical weight
of the cycle which plays the key role both in the classical and
stochastic trace formulas.

In diagrammatic language, we join all possible pairs of $\varphi$
vertices, each one giving a $\sigma^2 C$ propagator. Thus the first
diagram in (\ref{e:expdiag}) is expanded (\ref{e:V''},\ref{e:phi})
to
\[
\btrackBB{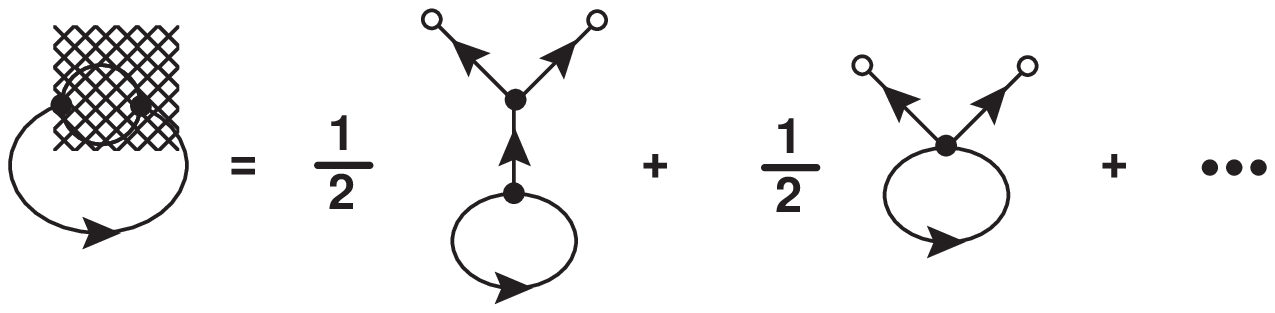}
\]
and then the $\varphi$ vertices joined to form two diagrams contributing
at order $\sigma^2$.  The full noise corrections of order $\sigma^2$ are
given by all connected two-loop diagrams:
\beq
W_{c,2}= {1\over 2} \btrack{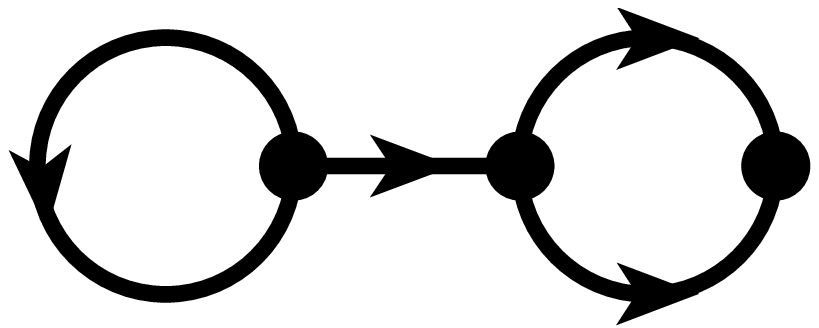}
	 + {1\over 2} \btrack{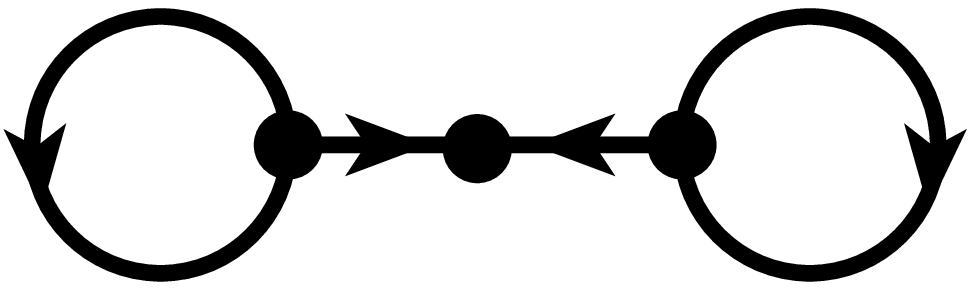}
	 + {1\over 2} \btrack{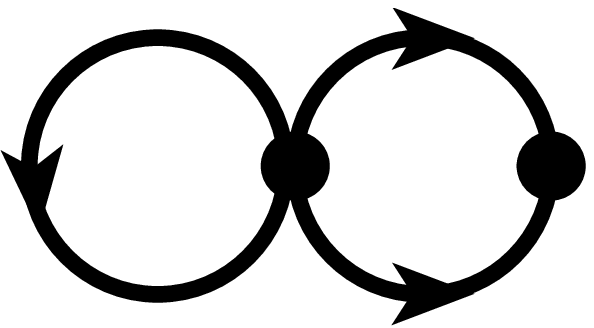}
	 + {1\over 2} \btrack{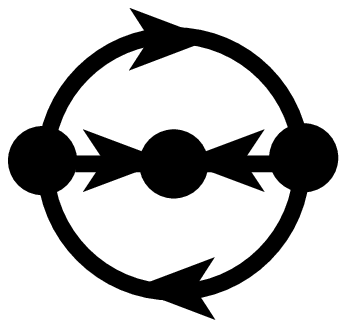}
\,.
\ee{a:sigSqGraphs}
Each diagram has a two-fold symmetry, hence all combinatorial
weights equal $1/2$.
Before writing down the final expression, we note that several
sub-diagrams may be simplified using (\ref{e:Prpgtr}).
These are
(no sum on $a$, $b$)
\bea
\btrack{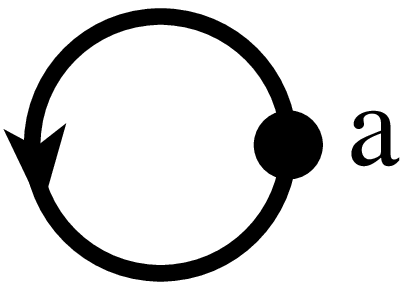}
&=&
  \Prpgtr{aa} \, = \, \frac{-\ExpaEig_c}{\ExpaEig_c-1} 
	\, \frac{1}{\Df{a}}
	% \qquad (\mbox{no sum on }a,\, b)
\\[\Fsize]
\nonumber
\btrack{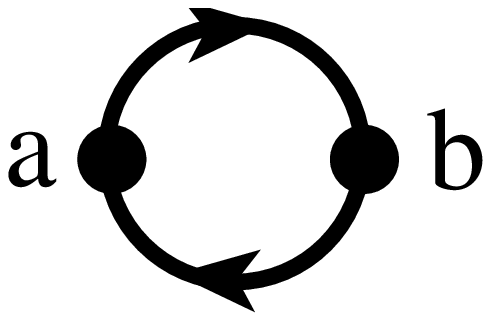}
&=&
\Prpgtr{ab}\Prpgtr{ba} \,=\,\frac{\ExpaEig_c}{(\ExpaEig_c-1)^2}
\frac{1+(\ExpaEig_c-1)\delta_{ab}}{\Df{a}\Df{b}}
\,.
\nonumber
\eea
Adding the terms we obtain the $\sigma^2$ contribution to
the trace:
\bea
W_{c,2}
	&=& {1 \over 2}
\left[\frac{\ExpaEig_c}{\ExpaEig_c-1}\sum_{a}
	\left(\frac{f^{''2}_a}{f^{'2}_a}-\frac{f^{'''}_a}{\Df{a}}\right)
\CrlMat{aa}
	\right.
	\ceq
	\left.
~~~+\frac{\ExpaEig^2_c+\ExpaEig_c}{(\ExpaEig_c-1)^2}
	\sum_{ab}\frac{f^{''}_a}{\Df{a}} \frac{f^{''}_b}{\Df{b}}\CrlMat{ab}
-\frac{\ExpaEig_c}{\ExpaEig_c-1}
	\sum_{ab}\frac{f^{''}_a}{\Df{a}} f^{''}_b\Prpgtr{ab}\CrlMat{bb}
\right]
\,.
\label{e:pp}
\eea
For an alternative approach to evaluating multiple derivatives, see
\refappe{s:recurs}.

\subsection{Repeats of prime cycles}
\label{s:Repeats}

In the deterministic case repeats of periodic orbits can be summed
up, and \fd s and \dzeta s written in terms of prime cycles
rather than periodic points.
In order to accomplish this for the stochastic case, 
we need to compute the trace for repeats of periodic orbits.

For $r$ repeats of a prime cycle $p$
we have $\cl{}=\cl{p}r$, $\ExpaEig_c=\ExpaEig_p^r$, where
$\ExpaEig_p$ is the stability of the prime cycle $p$.
Each index $a=0,\ldots,n$$-$$1$ is decomposed as
$a=\mod{a}+\bar{a}\cl{p}+\const{a}$ with
$\mod{a}=0,\ldots,$$\cl{p}$$-1$ and $\bar{a}=0,\ldots,r$$-$$1$.  $\const{a}$ is an
arbitrary starting point on the orbit which may be chosen independently
for each index.  $f$ and its
derivatives depend only on $\mod{a}$. 

The first sum in \refeq{e:pp} is
\[
\frac{1}{2}
\sum_{\mod{a}\bar{a}\mod{b}\bar{b}}
\frac{\ExpaEig_p^r}{(\ExpaEig_p^r-1)^3}
	\left(
		\frac{f^{''2}_{\mod{a}}}{f^{'2}_{\mod{a}}}
		-\frac{f^{'''}_{\mod{a}}}{\Df{\mod{a}}}
        \right)
     \ExpaEig_p^{2(\bar{a}-\bar{b})}
      \prod_{\mod{d}=\mod{b}+1}^{\mod{a}-1}f^{'2}_{\mod{d}}
\]
where $\bar{a}-\bar{b}$ is the number of full repeats of $p$ contained in
$\prod_{d=b+1}^{a-1}$; this is achieved by setting $\const{a}=b+1$.  The sums
over $\bar{a}$ and $\bar{b}$ are performed, leading to
\[
\frac{r}{2}\frac{\ExpaEig_p^{2r}-1}{\ExpaEig_p^2-1}
 \frac{\ExpaEig_p^r}{(\ExpaEig_p^r-1)^3}
  \sum_{\mod{a}\mod{b}}
    \left(
          \frac{f^{''2}_{\mod{a}}}{f^{'2}_{\mod{a}}}
	 -\frac{f^{'''}_{\mod{a}}}{\Df{\mod{a}}}
    \right)
\prod_{\mod{d}=\mod{b}+1}^{\mod{a}-1}f^{'2}_{\mod{d}}
\]
which is just a factor depending on $r$ and $\ExpaEig_p$ multiplied by the
sum for the single repeat of the prime cycle.

When the calculations are carried out for both of the other sums, some
rather unenlightening algebra leads to exactly the same prefactor;
we discuss this rather remarkable point and its generalization to higher
orders in detail in the sequel paper\rf{conjug}.   Combining all three
terms of (\ref{e:pp}) leads to an expression for the trace in terms of cycles:
\bea
\tr{z\Lnoise{} \over 1 - z\Lnoise{}} 
 &=& 
\sum_{n=1}^\infty z^n \sum_{x_c\inFix{n}} e^{W_c}
\continue
 &=&
 \sum_p\cl{p}\sum_{r=1}^{\infty}
\frac{z^{\cl{p}r}}{|\ExpaEig_p^r-1|}
\exp\left\{\frac{\sigma^2}{2} w_{p,2}
	  \frac{\ExpaEig_p^r(\ExpaEig_p^r+1)}{(\ExpaEig_p^r-1)^2} r
	  \right\}
\label{SdlptTraceForm}
\eea
up to order $\sigma^2$, where
\bea
w_{p,2}&=&\frac{\ExpaEig_p-1}{\ExpaEig_p+1}\sum_{a}\left(\frac{f^{''2}_a}{f^{'2}_a}
-\frac{f^{'''}_a}{\Df{a}}\right)\CrlMat{aa}
	\ceq
+\sum_{ab}\frac{f^{''}_a}{\Df{a}} \frac{f^{''}_b}{\Df{b}}\CrlMat{ab}
-\frac{\ExpaEig_p-1}{\ExpaEig_p+1}\sum_{ab}\frac{f^{''}_a}{\Df{a}}f^{''}_b
\Prpgtr{ab}\CrlMat{bb}
\label{e:PrimCycSigSq}
\eea
contains all the dependence on the higher derivatives along
the prime cycle $p$, with no dependence on the repetition number $r$.
To put it another way, if $p$ is a cycle, not necessarily prime, then
\begin{equation}
w_{p^r,2}=rw_{p,2}\;\;.\label{e:simplerepeats}
\end{equation}

Next, using the identity
\[
        {1+x \over (1-x)^3} = \sum_{k=0}^{\infty} (k+1)^2 x^k
\]
we rewrite the trace formula in a form in which 
repeats are resummed over by expanding the exponential in
(\ref{SdlptTraceForm}) to order $\sigma^2$, forming the sum over $k$,
and putting the result back in an exponential:
\bea
\tr{z\Lnoise{} \over 1 - z\Lnoise{}}
	&=&
\sum_p\cl{p}\sum_{k=0}^{\infty}\sum_{r=1}^{\infty}
\frac{z^{\cl{p}r}}{|\ExpaEig_p^r|\ExpaEig_p^{kr}}
e^{r\frac{\sigma^2}{2}(k+1)^2 w_{p,2}} \,+\,O(\sigma^4)
	\continue
	&=&
\sum_p\cl{p}\sum_{k=0}^{\infty}
	{t_{p,k} \over 1-t_{p,k}}
\,,
\label{NoiseResum}
\eea
where $t_{p,k}$ is the $k$-th local eigenvalue
\[
t_{p,k} = \frac{z^\cl{p}}{|\ExpaEig_p|\ExpaEig_p^k}
	  e^{\frac{\sigma^2}{2} (k+1)^2 w_{p,2}}
 \,+\,O(\sigma^4)
\]
This is the stochastic equivalent of the Gutzwiller trace formula
for the semiclassical case~\rf{gutbook}.

We sum over $r$ as usual\rf{QCcourse} to
obtain from \refeq{NoiseResum} a Selberg type product for the noisy {\Fd}
\beq
\det(1-z\Lnoise{})
=\prod_p\prod_{k=0}^{\infty}
\left(1-  t_{p,k}\right)
\,,
\ee{e:FredD}
valid to order $\sigma^2$.

We observe a crossover effect, since for higher order eigenvalues
(large $k$), eventually the argument of the exponential becomes of
order one, and further noise corrections are required.  
This is as it should be: the higher order eigenfunctions
have more detailed structure, are more quickly smeared by the noise,
and should decay faster.

\subsection{Fixed point}

The contribution from a fixed point (cycle of length one) is particularly
simple, as all the sums and products collapse to a single term, and
$\Df{}=\ExpaEig$.  We obtain 
\begin{equation}
w_{p,2}=\frac{1}{\ExpaEig(\ExpaEig+1)}\left\{
3\left(\frac{f^{''}}{\ExpaEig-1}\right)^2-\frac{f^{'''}}{\ExpaEig-1}\right\}
\end{equation}

If the map contains only a single isolated unstable fixed point,
we thus have an expression for the eigenvalues,
\[
\eigenvL_k(\sigma)=
\frac{1}{z_k(\sigma)}
 =\frac{1}{|\ExpaEig|\ExpaEig^k}
		e^{\frac{\sigma^2}{2} (k+1)^2 w_{p,2}}
\]
valid to order $\sigma^2$.  Note that depending on the sign of $w_{p,2}$,
small amounts of noise can either enhance or inhibit escape from the
fixed point.
Higher order terms for a fixed point are given in 
\refref{conjug}.

\section{Numerical tests}
\label{e:NumTests}

To test the above expressions for the trace, we have computed
the required derivatives for the 23 
prime cycles up to
length $\cl{} =6$ for the  quartic map
\beq
f(x)=20\left({1\over 2^4}-\left({1\over 2}-x\right)^4\right)
\,.
\ee{e:quartRep}
The choice of the map is motivated by requiring that
the system be simple (one-dimensional in this case),  
with non-trivial $f''$, $f'''$ (hence quartic),
with complete binary dynamics (hence a nice repeller),
and no diffraction and nonhyperbolic regions in the immediate
vicinity of the repeller (where the Gaussian
saddle points would be insufficient). 

\subsection{Evaluation of the determinant}
In this we follow the approach to computing escape rates originally
introduced by Kadanoff and Tang\rf{KT}.
The topological length truncated 
cycle expansions\rf{QCcourse} of \Fd\ \refeq{e:FredD} 
are obtained by writing the trace and determinant as power
series expansions in $z$ and $\sigma$,
\begin{eqnarray}
\tr{z\Lnoise{} \over 1 - z\Lnoise{}}
&=&\sum_{n=1}^{N}z^n(C_{n,0}+\sigma^2 C_{n,2})\\
\det(1-z\Lnoise{})&=&1-\sum_{n=1}^{N}z^n(c_{n,0}+\sigma^2 c_{n,2})
\end{eqnarray}
Here, the $C$ coefficients come from \refeq{SdlptTraceForm}, and the
$c$ coefficients are obtained by equating coefficients in
\begin{equation}
\det(1-z\Lnoise{})\,\tr{z\Lnoise{} \over 1 - z\Lnoise{}}
=-z\frac{d}{dz}\det(1-z\Lnoise{})
\end{equation}
following from the identity $\ln\det M=\tr\ln M$.  The solution is found
recursively as
\begin{eqnarray}
c_{n,0}&=&\frac{1}{n}\left[C_{n,0}-C_{n-1,0}c_{1,0}-\ldots-C_{1,0}c_{n-1,0}
\right]\\
c_{n,2}&=&\frac{1}{n}\left[C_{n,2}-(C_{n-1,0}c_{1,2}+C_{n-1,2}c_{1,0})
-\right.\nonumber\\& &\left.\ldots-(C_{1,0}c_{n-1,2}+C_{1,2}c_{n-1,0})\right]
\end{eqnarray}
From the $c_{n,0}$ coefficients we construct the deterministic Fredholm
determinant, from which the deterministic eigenvalue $\eigenvL_0$ is
found using Newton's method on the characteristic equation for 
$\Lnoise{}$ at $\sigma=0$:
\begin{equation}
1-\sum_{n=1}^{N}\eigenvL_0^{-1}c_{n,0}=0
\end{equation}
The $\sigma^2$ correction to the eigenvalue is found from the $\sigma^2$
terms in the characteristic equation, and comes to
\begin{equation}
\eigenvL_2=-\frac{\sum_{n=1}^{N}\eigenvL_0^{-n}c_{n,2}}
{\sum_{n=1}^{N}n\eigenvL_0^{-n-1}c_{n,0}}
\end{equation}
The leading eigenvalue $\eigenvL_0$ for the deterministic 
(noiseless) map and the coefficient of the
$\sigma^2$ correction $\eigenvL_{0,2}$,
shown in \reftab{t:Fred}, demonstrate the superexponential
convergence with $n$ of \Fd, as expected 
for nice hyperbolic dynamical systems.  

The escape rate $\gamma$ of the repeller
is calculated directly from the eigenvalue,
\[
\gamma(\sigma)=-\ln\eigenvL(\sigma)
\]
and hence
\begin{eqnarray*}
\gamma_0&=&-\ln\eigenvL_0\\
\gamma_2&=&-\frac{\eigenvL_2}{\eigenvL_0}
\end{eqnarray*}
We have also directly tested
the repeat formula (\ref{e:simplerepeats}) for our cycle set.

\begin{table}
\begin{tabular}{cll}
\hline
$n$&$\eigenvL_0$&$\eigenvL_{0,2}$\\
\hline 
1&    0.308            &     0.42            \\
2&    0.37140          &     1.422           \\
3&    0.3711096        &     1.43555         \\
4&    0.371110995255   &     1.435811262     \\
5&    0.371110995234863&     1.43581124819737\\
6&    0.371110995234863&     1.43581124819749\\
\hline
\end{tabular}
\caption{
Significant digits of the leading deterministic 
eigenvalue and its $\sigma^2$ coefficient,
calculated from the \Fd\ as function of the cycle truncation length $n$.
Note the superexponential convergence of both 
$\eigenvL_0$ and $\eigenvL_{0,2}$ ($n=6$ result is limited by the 
machine precision).
}
\label{t:Fred}
\end{table}

\subsection{Discretized eigenfunction}
In order to check the perturbative calculation we
evaluate the eigenvalue numerically as a function of
$\sigma$ by treating the evolution operator as a matrix acting on a
discretized eigenfunction.  That is, we approximate $\cal L$ by a matrix
${\cal L}_{y_cx_c}$ where $(x_c,y_c)$ is the center of a square in the $x$-$y$
plane of small but finite size $\epsilon$ (with upper limits $(x_t,y_t)$
and lower limits $(x_b,y_b)$).  The matrix element is obtained by assuming the
distribution $\rho$ is constant across this small square:
\[
{\cal L}_{y_cx_c}=\epsilon^{-1}\int_{x_b}^{x_t}\int_{y_b}^{y_t}
\delta_\sigma(y-f(x))dx dy
\,.
\]
The integral may be approximated
using just a few evaluations of the kernel with errors of a similar order
to those due to the variation in $\rho$.  
For example
\[
{\cal L}_{y_cx_c}=\frac{1}{16}(L_{tt}+2L_{tc}+L_{tb}+2L_{ct}+4L_{cc}+2L_{cb}+
L_{bt}+2L_{bc}+L_{bb})
\]
with $L_{tc}=\delta_\sigma(y_t-f(x_c))$,
{\em etc.} requires only four evaluations per
square since the boundary points belong to more than one square.
For the map at hand, the discretized \evOper\ leads
to six digit accuracy in the escape rate for values of $\sigma$ as low
as $3\epsilon$.

When $\sigma$ is small the matrix is very sparse, a fact which can be used
to speed up the calculation.  The leading eigenvalue is obtained by repeatedly
evolving and rescaling an arbitrary smooth initial distribution, which
then approaches the leading eigenfunction.  In our case the map is
expanding on the neighborhood of the asymptotic repeller,
so the eigenfunctions are smooth, and
the discretization procedure is stable.  
The numerical eigenfunctions for two values of $\sigma$ is shown in
\reffig{f:eigenf}.  
The eigenfunction peaks at the critical point of the
map, but this has no detectable
effect on the eigenvalue, as subsequent iterations
send the points around the peak towards $-\infty$,
away from the repeller $x\in [0,1]$.  As the flow
is conserved only on the infinite interval
$x\in(-\infty,\infty)$, no normalizable
eigenfunction exists.  On any finite interval, however, the escape rate
is non-zero. If no point outside the interval can return (ignoring
the exponentially small tunneling probabilities), 
that is, if the interval encloses the
repeller and the neighboring
region around it determined by the magnitude of the
noise, the escape rate and the leading eigenfunction
(up to normalization) are
independent of the interval chosen.

Subtracting the perturbative analytic terms from the numerically 
computed $\eigenvL_0(\sigma)$,
\beq
\eigenvL_0(\sigma) - 0.371110995234863-1.43581124819749\sigma^2
\sim O(\sigma^4)
\ee{e:semiAnEst}
we compare with the numerically computed $\eigenvL_0(\sigma)$
in \reffig{f:eigenv},
and estimate the next term to be approximately $38\sigma^4$.

\FIG{
\includegraphics[width=0.85\textwidth]{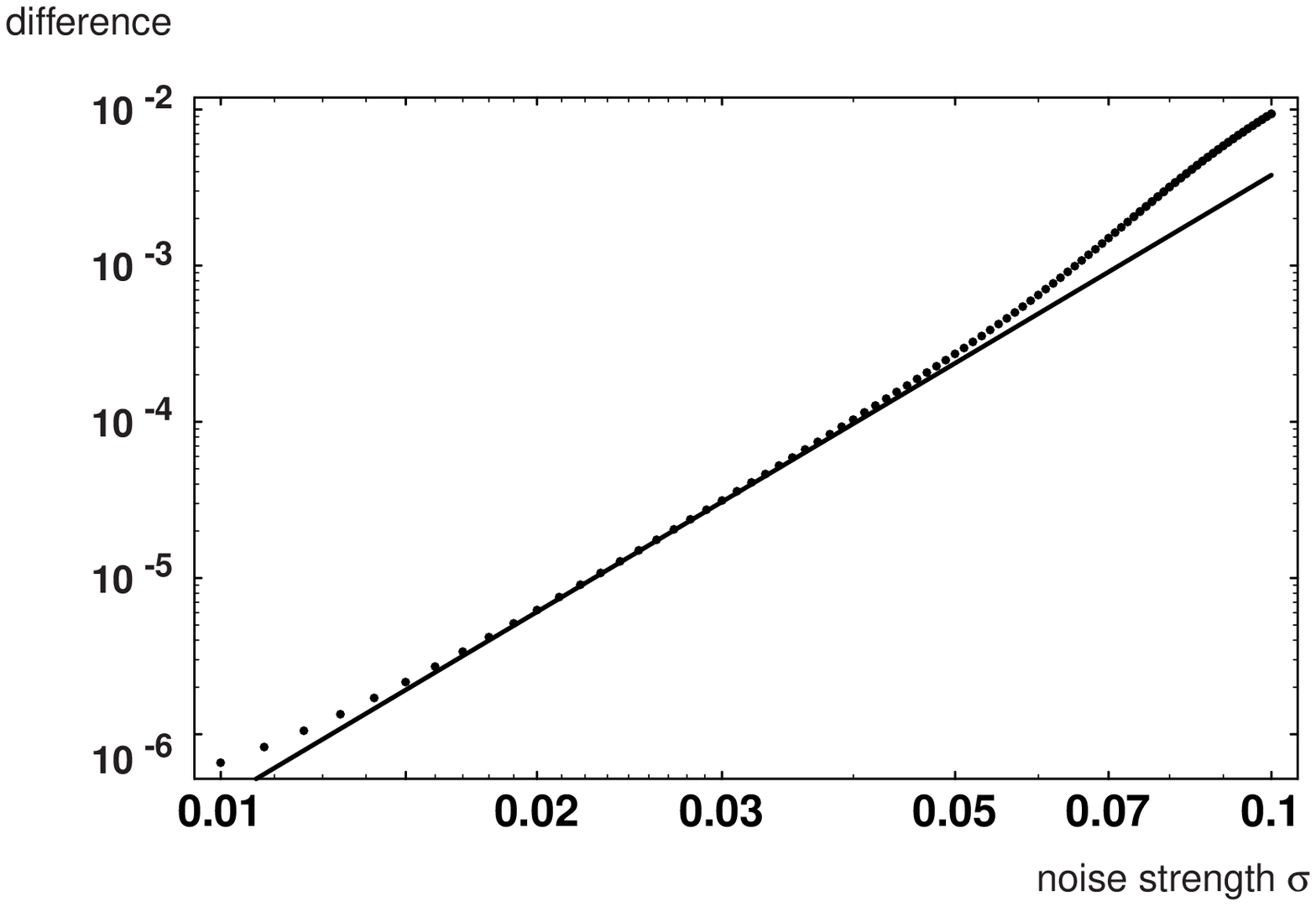}
}
{}{
The deviation of the
analytic estimate \refeq{e:semiAnEst} from the
numerically computed $\eigenvL_0(\sigma)$ for a range of
values of the noise strength $\sigma$ (points), {\em vs.} the
conjectured remainder, $38\sigma^4$ (solid line).  For small
$\sigma$ the errors are due to the finiteness of the grid, and for
larger $\sigma$ the deviation is due to the neglected
higher order contributions.
}
{f:eigenv}

\FIG{
\includegraphics[width=0.85\textwidth]{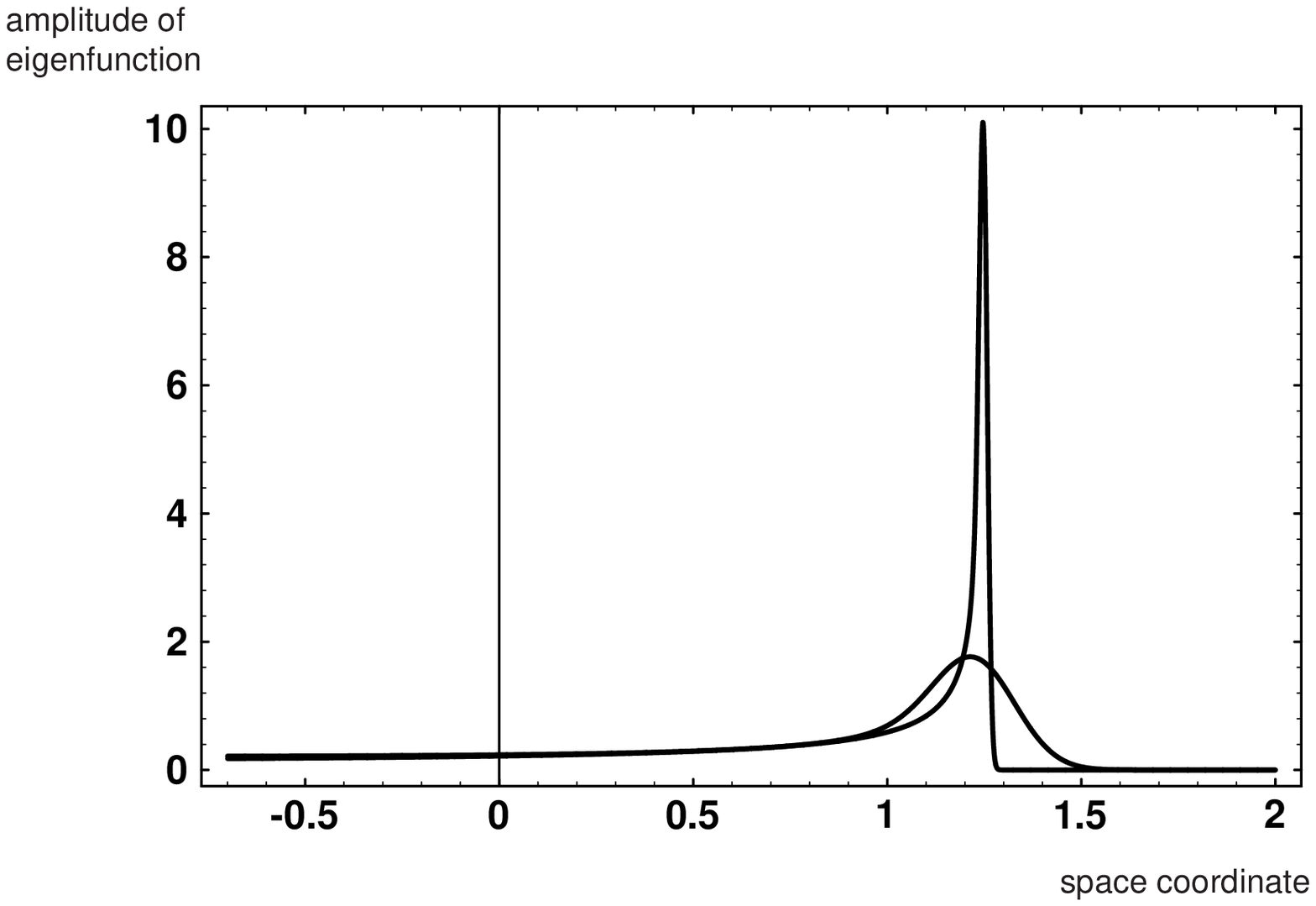}
}
{}{
The eigenfunction for $\sigma=0.01$ (sharp peak) and $\sigma=0.1$ (smoother).
}
{f:eigenf}

\section{Summary and outlook}

We have formulated weak noise perturbation theory for noisy maps in
terms of periodic orbits of the deterministic system, expanding to
order $\sigma^2$ explicitly, resummed repeats of prime cycles, and
tested the results numerically.  From here, there are many possible
generalizations and future directions.

In the sequel paper\rf{conjug} we shall recast the remarkable
resummation of repeats (sect.~\ref{s:Repeats}) in a more general framework
applicable to all orders of the expansion.  It seems from the
numerical section that the coefficients of powers of $\sigma$ are
growing very rapidly. The expansion is expected to be asymptotic.

Evaluation of expectation values\rf{QCcourse} on a stochastic flow 
requires replacing the \FPoper\ \refeq{Lnoise} by the generalized
evolution operator
\beq
\Lnoise{} =\delta_\sigma(y-f(x)) \,  e^{\beta \cdot \obser(x)}
\,.
\ee{LnoiseGen}
The same general perturbation theory applies, but now an
observable $\obser(x)$  contributes an extra set of
interaction vertices to $S_I(\field)$ in \refeq{PropIntr1}.
Similarly, the addition of more dimensions and/or non-Gaussian weak
noise can be treated by modifying the propagators and adding new
vertices.

While for deterministic flows it is appropriate to replace a flow
by a return map on a Poincar\'e section of the flow, it is not
clear that this is appropriate for stochastic flows; a noise
that is ``white'' on the Langevin equation level is ``colored''
when integrated to a Poincar\'e section return, and it might have
memory of the trajectory that a noisy iterated mapping cannot mimic.

The noise in general has a different structure than the deterministic
equations of motion; it typically breaks whatever
symmetries the classical flow might have, unless clever precautions
are taken to ensure that the noise
respects the symmetry\rf{CK}.
This situation is familiar
from Quantum Mechanics, where quantization and canonical transformations
do not commute. 

Our saddlepoint approximation to the spectrum of the exact
\evOper\ receives perturbative contributions from
all cycles, no matter how long.
However, the noise causes the physical system to effectively
lose memory at a rate depending on the region of phase space, so it might
be possible to obtain accurate averages by replacing the
\evOper\ by effective finite memory, finite Markov partition
transfer matrices.

Such studies might enable us
to understand the range of applicability of the ``semi-classical''
theory in greater detail than for the single
cut-off time proposed in the case of semiclassical quantization
 by Berry and Keating\rf{BK90}.   Different regions
of phase space are dominated by different time scales, and the program of
periodic orbit theory allows us to use the dynamics itself,
encoded in the properties
of cycles, to determine at what point classical behavior is modified by
semiclassical or noise corrections.

As in the semiclassical case, the saddlepoint approximation causes
the multiplicative structure of the evolution operators to be lost, and
one might consider extended formulation of \refref{CV93} to 
improve the analyticity of the \fd s.  Finally, non-analytic points in
the dynamics will lead to diffraction effects which are of different
orders in $\sigma$, for example, the escape rate of the map $4x(1-x)$
which has a quadratic maximum at the boundary of the deterministic
repeller is of order $\sqrt{\sigma}$\rf{reimann1}.

%%-------------------------------------------------------
%%-----   Appendices
%%-------------------------------------------------------
\appendix

\section{Appendix: Recursive evaluation of derivatives}
\label{s:recurs}

The derivatives of $x_n = f^n(x)$
\beq
        x_n' = {d x_n \over dx}
                \,, \quad x_n'' = {d^2 x_n \over dx^2}
                \,, \quad x_n''' = {d^3 x_n \over dx^3}
        \,, \dots \,,
\eeq
with initial values
\beq
        x_0 = x\,, \quad {d x_0 \over dx} = 1
                \,, \quad{d^2 x_0 \over dx^2} = 0
                \,, \quad {d^3 x_0 \over dx^3} = 0
        \,, \dots \,,
\eeq
can be computed recursively by
\bea
        x_{n+1}' &=& f'(x_{n}) x_{n}'
                          = \prod_{k=0}^{n} f'(x_k)
                         \continue
        x_{n+1}'' &=&
                        f''(x_{n}) \left(x_{n}'\right)^2 + f'(x_{n}) x_{n}''
                         = x_{n+1}' \sum_{k=0}^{n}
                           { f''(x_k) \over f'(x_k)} x_k'
                         \continue
        x_{n+1}''' &=&
                        f'''(x_{n}) \left(x_{n}'\right)^3
                        +3 f''(x_{n}) x_{n}' x_{n}''
                        + f'(x_{n}) x_{n}'''
                                \continue
                         &=& x_{n+1}' \sum_{k=0}^{n}
                   { f'''(x_k) \over f'(x_k)} \left( x_k'\right)^2
			+3x_{n+1}' \sum_{0\leq j<k\leq n}
			\frac{f''(x_j)f''(x_k)}{f'(x_j)f'(x_k)}
			x_j'x_k'
                                \continue
        x_{n+1}'''' &=&
                        f''''(x_{n}) \left(x_{n}'\right)^4
                        +6 f'''(x_{n}) x_{n}'' (x_{n}')^2
                                \ceq
                        +4 f''(x_{n}) x_{n}''' x_{n}'
                        +3 f''(x_{n}) (x_{n}'')^2
                        + f'(x_{n}) x_{n}''''
        \,,
\eea
%\PC{recheck the last $x_{n+1}'''$ (this formula is wrong),
%incorporate in what follows...}
%\PC{Perhaps add $x_{n+1}^{(5)}$ recursion here.}
$x_{n+1}'$ in the above has form of a propagator,
$ f''(x_k) / f'(x_k) $ 3-vertex,
$x_{n+1}'''$ gets contribution from a 4-vertex diagram plus three
4-leg diagrams with two 3-vertices, {\em etc.}. In another words,
this iteration of the ``transport equations'' generates
the Feynman diagram expansion.

%REFERENCES -------------------------------------------------------
\renewcommand{\baselinestretch} {1}
{\small
\bibliography{cites}
\bibliographystyle{unsrt}
} % end of small

\end{document}